# ArcadeDB in Python: An In-Process Multi-Model Database

**Taewoon Kim**[1]

[1]HumemAI

## Abstract

Some Python workflows need transactional reads and writes over documents, traversals over a graph, and similarity search over vector embeddings. In practice these are split across separate systems (a document store, a graph engine, and a vector index), and the data is copied between them and kept consistent by application code. ArcadeDB is an open-source multi-model database engine, written in Java, that supports all three access patterns and can be embedded in a host process. We present `arcadedb-embedded`, a Python package that runs ArcadeDB inside the Python process via JPype, using a Java Runtime Environment bundled in the wheel. One `pip install` gives a database with documents, a property graph, and vector search, and needs no Java installation. Query results can be returned as Python lists, NumPy arrays, pandas DataFrames, or Arrow tables. We demonstrate a hybrid retrieval workflow that chains vector search, SQL filtering, and graph traversal over one dataset. We benchmark `arcadedb-embedded` against SQLite, DuckDB, LadybugDB, and Chroma, and report its costs.

## 1. Introduction

A question-answering system over a Q&A site stores posts and their metadata as documents (queried with SQL, much like rows of a table, but without a fixed schema), models who-answered-what or what-cites-what as a graph, and retrieves passages by semantic similarity over vectors, all over the same corpus. Python is popular for these workloads, but no single Python-embeddable data store covers all three access patterns. The usual response is to assemble a stack of SQLite or DuckDB for documents, NetworkX or an embedded graph database such as LadybugDB for the graph, and a vector index such as FAISS, hnswlib, or Chroma for vectors [1], [2], [3], [4], [5], [6], [7].

A stack of specialists is a reasonable default, and D. Datta [8] makes the case for assembling one from interoperable Python libraries. It has costs when all three access patterns apply to the same data. Each system holds its own copy, with its own consistency model, so every write has to reach all of them. A query that spans models passes identifiers from one library to another. Reproducing an experiment means pinning several packages instead of one.

This paper describes the alternative, one embedded multi-model database that holds documents, a property graph, and vector indexes. That database is ArcadeDB [9], an Apache-2.0 Java application that keeps all three models in one store (referred to as “the engine” throughout the rest of the paper). Documents are queried with its SQL dialect, the graph with OpenCypher, and vectors through a graph index of the Hierarchical Navigable Small World (HNSW) family. Transactional (online transaction processing, OLTP) and analytical (online analytical processing, OLAP) queries run over the same data. ArcadeDB is ACID (atomic, consistent, isolated, durable). Its Graph Analytical Views (GAV), an ArcadeDB-specific structure, speed up graph analytics while the base graph stays transactional.

**Correspondence to**
Taewoon Kim
taewoon@humem.ai

`arcadedb-embedded` binds the Java engine through JPype [10], [11]. The paper contains benchmarks comparing `arcadedb-embedded` to the specialists.

Concretely, this paper contributes:

1. **An in-process multi-model database for Python.** Each wheel bundles a Java Runtime Environment (JRE) for its platform, from which the binding starts a Java Virtual Machine (JVM) inside the Python process via JPype. The binding covers JVM lifecycle, transactions, and NumPy/pandas/Arrow interop. Wheels exist for four platforms and five Python versions, and the user never installs Java.
2. **A hybrid workflow.** A single retrieval pipeline composes vector search, SQL filtering and graph traversal over one dataset in one process. A capability matrix shows that no single Python-embeddable alternative expresses this composition.
3. **A benchmark against embedded specialists.** A reproducible benchmark suite compares the engine, from Python, against embedded specialists on a real-world Q&A corpus, across transactional, analytical, and vector workloads. We report which models each side wins and what the unified engine costs in memory, startup, latency tails, disk, and result transport into Python.

## 2. Background

**Embedded versus server databases.** A server database, such as PostgreSQL with pgvector [12], Neo4j [13], or Milvus [14], runs as its own process, and Python reaches it over a socket. An embedded database is a full database engine delivered as a library. It runs inside the host process and persists to ordinary files, which fits local, single-machine work. This paper is about that setting, so from here on it compares `arcadedb-embedded` only with other Python-embeddable stores.

**Embedded data stores in Python.** SQLite is the standard choice for documents and transactions [2]. DuckDB brought a columnar, analytics-first embedded engine to Python [1], [15]. For graphs, NetworkX [3] provides in-memory graph analysis. LadybugDB [4] is an embedded graph database with a Cypher interface and a vector index. For vectors, hnswlib provides the reference HNSW implementation [6], [16], FAISS provides exact and approximate similarity search [5], and Chroma packages HNSW behind a Python API [7]. None of them provides all three models in one embedded engine (Table 1).

**Table 1.** *Capability matrix for the Python-embeddable data stores compared here; all five run in the host process (✓ native, ∼ partial or through an official extension, × absent). "All three" means documents with SQL, a property graph, and vector search. DuckDB's HNSW index is an experimental core extension; ArcadeDB's index is JVector [17] (Vamana graphs [18] with an HNSW-style hierarchy). SQLite has vector search only through extensions, and neither of them is a graph index. SQLite and DuckDB are relational, and LadybugDB is a graph database whose nodes live in typed node tables. Chroma stores each entry as a text document with a metadata dict. ArcadeDB's SQL has `GROUP BY` and aggregates but no joins and no `OVER`/`PARTITION BY` window clause.*

| Capability | SQLite | DuckDB | LadybugDB | Chroma | ArcadeDB |
|---|---|---|---|---|---|
| Documents | ✓ | ✓ | ∼ | ∼ | ✓ |
| Transactional SQL (OLTP) | ✓ | ∼ | × | × | ✓ |
| Analytical SQL (OLAP) | ∼ | ✓ | × | × | ∼ |
| Property graph + traversal | × | × | ✓ | × | ✓ |
| Cypher | × | × | ✓ | × | ✓ |
| Graph approximate-nearest-neighbor index (HNSW family) | × | ∼ | ✓ | ✓ | ✓ |
| All three in one engine | × | × | × | × | ✓ |

## 3. Architecture

`arcadedb-embedded` runs the ArcadeDB engine *inside* the Python interpreter process. Opening a database is a function call, and queries are method calls that return Python objects.

**In-process JVM via JPype.** On first use, the binding starts an embedded JVM inside the Python process using JPype [10]. JVM settings like heap size are configured directly from Python, and CPython and the JVM share a single address space, so data can be passed around without needing network calls or inter-process serialization. The engine's Java objects are exposed through a Pythonic layer, so applications never need to interact with JPype directly.

```
import arcadedb_embedded as arcadedb

# One in-process engine; persists to a local directory.
with arcadedb.create_database("my_db", jvm_kwargs={"heap_size": "4g"}) as db:
    db.command("sql", "CREATE DOCUMENT TYPE Post")
    db.command("sql", "CREATE PROPERTY Post.id LONG")
    db.command("sql", "CREATE INDEX ON Post (id) UNIQUE_HASH")
    with db.transaction():
        db.command("sql", "INSERT INTO Post SET id = :id, score = :s",
                   {"id": 1, "s": 42})
    rows = db.query("sql", "SELECT id, score FROM Post WHERE id = 1").to_list()
```

**One object for all three models.** The same `db` object queries documents (SQL), the property graph (OpenCypher), and vectors (a graph index, searched with the `vectorNeighbors()` function). The first argument of `query()` selects the language. All three models share one transaction scope and one set of identifiers, so the hybrid workflow below can pass results from one model into the next without leaving the process.

**NumPy, pandas, and Arrow.** Embeddings are passed as `float32` NumPy arrays through `arcadedb.to_java_float_array(vec)`, which hands the buffer to the engine without a Python-side element-by-element copy. Results come back through four methods on the result set. `to_list()` gives a list of dicts, `to_dataframe()` a `pandas.DataFrame`, `to_columns()` a dict of NumPy arrays, and `to_arrow()` a `pyarrow.Table`.

```
# db and the Post type as above; a result set is read once, so each output format
# needs its own query
df = db.query("sql", "SELECT id, score FROM Post").to_dataframe()   # pandas.DataFrame
cols = db.query("sql", "SELECT id, score FROM Post").to_columns()   # {"id": ndarray, ...}
table = db.query("sql", "SELECT id, score FROM Post").to_arrow()    # pyarrow.Table;
                                                       # nullable INTEGER stays int64
```

pandas and pyarrow are optional dependencies, imported only when the corresponding method is called. NumPy has no missing value for integers, so with `to_columns()` a nullable `INTEGER` column becomes `float64` with `NaN` holes, which loses the type and any precision above $2^{53}$. `to_arrow()` reads the same buffer and keeps its null bitmap, so the column stays `int64`. Both read one packed columnar buffer from the engine. Section 6 measures the four export paths.

**Cross-platform packaging.** Bringing an engine written in another language into Python usually means compiling a native C/C++ extension, with tooling such as scikit-build-core [19]. `arcadedb-embedded` ships a JRE instead. Java bytecode is platform-agnostic, so the engine's `.jar` files are identical everywhere and the only platform-specific part is the JRE, which is trimmed with `jlink` [20] to the modules the engine needs. The build produces wheels for Linux x86-64, Linux ARM64, macOS Apple Silicon, and Windows x86-64, for Python 3.10 to 3.14, twenty wheels in total. `pip install arcadedb-embedded` needs no Java installation and no `JAVA_HOME` (Figure 1). Each wheel is about 70 MB, roughly four times a NumPy wheel, downloaded once at install.

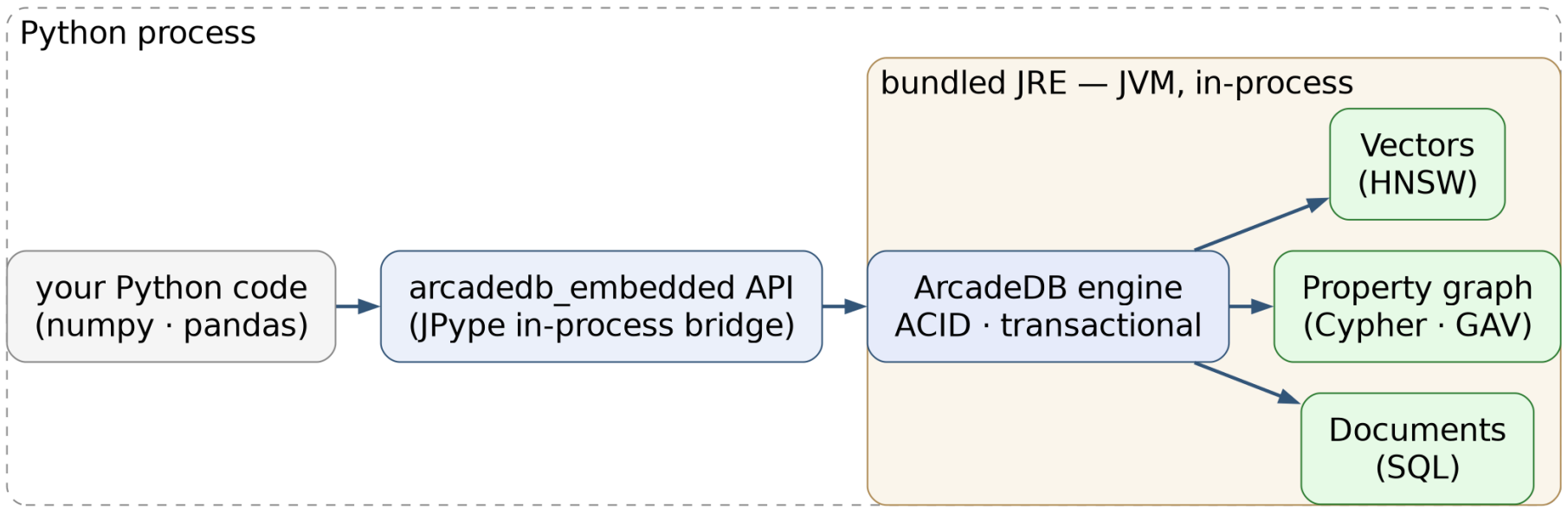

**Figure 1.** *The `arcadedb-embedded` architecture. A single Python process hosts a JVM (via JPype) running the ArcadeDB engine over one local, file-backed database. The Python API exposes documents/SQL, a property graph (OpenCypher), and vector search through one `db` object, with NumPy/pandas interop. The wheel bundles a per-platform JRE, so installation is a plain `pip install` and the user never installs Java.*

## 4. Documents, graph, and vectors in one process

This section first shows how data gets in, then each model with its own query language, and finally one workflow that uses all three. The data is the Cross Validated (`stats.stackexchange.com`) Stack Exchange dump, a statistics and machine-learning Q&A site (CC BY-SA [21]). Questions and answers are documents, who-asked and who-answered form a graph, and each question text has an embedding. The benchmark suite linked at the end of the paper rebuilds this corpus with one command and runs the whole section end to end.

### *4.1. Ingestion*

Data enters all three models as Python objects: `insert_many()` takes a list of dicts for documents, `graph_batch()` takes lists of dicts and ids for vertices and edges, and embeddings go in as `float32` arrays. The examples below start from the Parquet files the benchmark suite prepares and read them with pandas, selecting only the columns a type needs, because ArcadeDB has no Parquet or Arrow reader. Loading costs more than in the specialists. On the Cross Validated corpus, ArcadeDB needs about 17.5 s for the 426k posts and about 27 s for the graph. SQLite and DuckDB read the same Parquet files natively in about 0.3 s, and LadybugDB bulk-loads the graph with `COPY` in about 0.6 s (Table 2, Table 3). Most of the load time goes to the unique hash index on `id` that the point-lookup workload relies on.

### *4.2. Documents*

ArcadeDB's documents are created and queried with SQL, so a type of documents is used much like a table in SQLite or DuckDB, and writes run inside ACID transactions. The snippet in Architecture creates a type and a `UNIQUE_HASH` index for point lookups, then runs a parameterized `INSERT` inside `db.transaction()`. Documents can be schema-flexible or have typed properties for indexing, and both kinds are stored in the same database as the graph and vectors below.

Loading a corpus row by row pays a Python-to-Java call per row, so bulk ingestion goes through `insert_many()`, which serializes a batch to one string and loops over it on the Java side. Only the columns the type needs are read into Python:

```python
import pandas as pd

data = "datasets/prepared/stats.stackexchange.com"   # written by the benchmark suite's prepare step

# read only the columns the type needs; the post body stays in the file
posts = pd.read_parquet(f"{data}/posts.parquet", columns=["id", "score", "title"])
db.insert_many("Post", posts.to_dict("records"), commit_every=10_000)
```

### 4.3. *Property graph*

Vertex and edge types are declared with SQL. OpenCypher can create them implicitly on first write, as in Neo4j, and can add indexes and uniqueness constraints, but the typed float-array property and the vector index below need SQL, so we declare everything there. Bulk loading goes through a `graph_batch()` context manager. `create_vertices()` and `new_edges()` each cross into the JVM once for a whole list and the batch commits on flush, instead of one crossing and one transaction per element. `batch_size` is the number of edges buffered before an automatic flush, and `create_vertices()` returns the ids, so edges reuse them instead of querying them back. Graph queries then use OpenCypher through the same `db` object:

```python
import pandas as pd

db.command("sql", "CREATE VERTEX TYPE Question")
db.command("sql", "CREATE PROPERTY Question.embedding ARRAY_OF_FLOATS")
db.command("sql", "CREATE VERTEX TYPE Answer")
db.command("sql", "CREATE VERTEX TYPE Userx")
db.command("sql", "CREATE EDGE TYPE HAS_ANSWER")
db.command("sql", "CREATE EDGE TYPE AUTHORED_BY")

data = "datasets/prepared/stats.stackexchange.com"   # written by the benchmark suite's prepare step

# read only the columns the types need; the post body stays in the file
posts = pd.read_parquet(f"{data}/posts.parquet",
                        columns=["id", "post_type", "owner_user_id", "score", "title"])
questions = posts[posts.post_type == 1]                # embeddings has one row per question
answers = posts[posts.post_type == 2].dropna(subset=["owner_user_id"])
users = pd.read_parquet(f"{data}/users.parquet", columns=["id", "reputation"])
links = pd.read_parquet(f"{data}/edges_answers.parquet")   # question_id, answer_id

with db.graph_batch(batch_size=100_000) as batch:
    # create_vertices() returns one RID per row, so edges reuse them instead of querying
    qrid = dict(zip(questions.id, batch.create_vertices("Question", [
        {"id": i, "score": s, "title": t, "embedding": e.tolist()}
        for i, s, t, e in zip(questions.id, questions.score, questions.title, embeddings)])))
    arid = dict(zip(answers.id, batch.create_vertices(
        "Answer", answers[["id", "score"]].to_dict("records"))))
    urid = dict(zip(users.id, batch.create_vertices("Userx", users.to_dict("records"))))
    batch.new_edges([qrid[q] for q in links.question_id], "HAS_ANSWER",
                    [arid[a] for a in links.answer_id])       # one call per edge type
    batch.new_edges([arid[a] for a in answers.id], "AUTHORED_BY",
                    [urid[int(u)] for u in answers.owner_user_id])

db.query("opencypher",
    "MATCH (q:Question)-[:HAS_ANSWER]->(a:Answer)-[:AUTHORED_BY]->(u:Userx) "
    "WHERE q.id = $qid RETURN a.id, a.score, u.reputation "
    "ORDER BY a.score DESC",
    {"qid": 3}).to_list()          # Cypher binds $name from a dict; SQL binds :name or ?
```

For graph *analytics* over the same data, a GAV is created (`CREATE GRAPH ANALYTICAL VIEW ... VERTEX TYPES (...) EDGE TYPES (...)`). The build runs in the background, so the statement returns before the GAV is usable and `SELECT status FROM schema:graphAnalyticalViews` is read until it reports `READY`. The GAV is an in-memory copy of the selected vertex and edge types, laid out for fast traversal, and it is kept up to date as the graph changes.

## 4.4. Vectors

Embeddings are stored as a float-array property and indexed with an `LSM_VECTOR` index. The index is a JVector graph index [17], built from Vamana graphs [18] in a multi-layer hierarchy comparable to HNSW. Search is a SQL function over the index. Parameters (`dimensions`, `similarity`, `maxConnections`, `beamWidth` = `ef_construction`) are set on the index, and the query supplies `ef_search`. `maxConnections` is the per-layer out-degree, while hnswlib's $M$ is half of it, because hnswlib builds its base layer at degree $2M$ and upper layers at $M$. Setting both to the same number compares a half-degree ArcadeDB graph against a full-degree hnswlib one, so we set `maxConnections` $= 2M = 32$:

```
import numpy as np

db.command("sql", "CREATE INDEX ON Question (embedding) LSM_VECTOR "
    'METADATA { "dimensions": 384, "similarity": "COSINE", '
    '"maxConnections": 32, "beamWidth": 100 }')

# embeddings: one 384-d float32 row per Question, built by the benchmark suite linked
# at the end of the paper. vectorNeighbors(index, query vector, k, ef_search)
seed = np.asarray(embeddings[0], dtype=np.float32)
hits = db.query("sql",
    "SELECT id, score, distance FROM "
    "(SELECT expand(vectorNeighbors(?, ?, ?, ?))) ORDER BY distance",
    "Question[embedding]", arcadedb.to_java_float_array(seed), 200, 100).to_list()
```

## 4.5. The hybrid workflow

Given a popular "seed" question, we answer one query: *find questions similar to this one, keep the well-scored ones, and return their best answers with the reputation of who wrote them.* The steps are a vector search, a SQL filter, and a Cypher traversal (Figure 2). They pass 200 vector candidates to the SQL filter, 50 that pass the filter to the graph traversal, and return the top 10 answers. The first two steps are kept apart so that each stage can be timed separately. An application does not have to split them, because `vectorNeighbors()` returns whole documents, so the filter can sit on the vector search in one statement, `SELECT id, title, score FROM (SELECT expand(vectorNeighbors(...))) WHERE score >= 1`, and only the graph traversal then needs a second call.

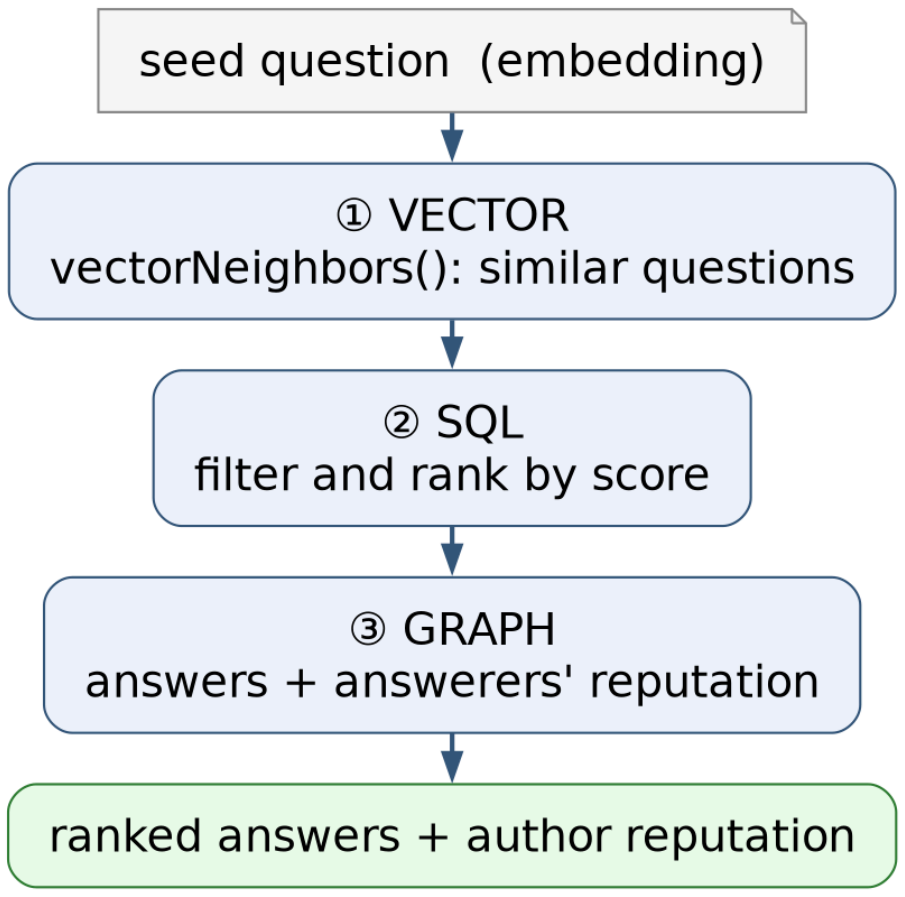

**Figure 2.** *The hybrid retrieval workflow. A seed question's embedding drives a vector search, the resulting ids flow into a SQL filter, and the surviving ids flow into a graph traversal that reaches answers and their authors. All three steps run against one database, with identifiers passing directly between steps.*

```
# 1) VECTOR: questions semantically similar to a seed
cands = db.query("sql",
    "SELECT id, score, distance FROM "
    "(SELECT expand(vectorNeighbors(?, ?, ?, ?))) ORDER BY distance",
    "Question[embedding]", arcadedb.to_java_float_array(seed), 200, 100).to_list()

# 2) SQL: keep well-scored candidates (ids flow straight from step 1)
ids = [int(c["id"]) for c in cands]
filt = db.query("sql",
    "SELECT id, title, score FROM Question WHERE id IN :ids "
    "AND score >= 1 ORDER BY score DESC LIMIT 50", {"ids": ids}).to_list()

# 3) CYPHER: traverse to answers + answerers' reputation
fids = [int(r["id"]) for r in filt]
hits = db.query("opencypher",
    "MATCH (q:Question)-[:HAS_ANSWER]->(ans:Answer)"
    "-[:AUTHORED_BY]->(usr:Userx) "
    "WHERE q.id IN $fids "
    "RETURN q.id AS qid, ans.id AS aid, ans.score AS ascore, "
    "usr.reputation AS rep ORDER BY ascore DESC LIMIT 10", {"fids": fids}).to_list()
```

The workflow runs on the part of the Cross Validated corpus its query needs: the 214k questions, their 209k answers, and the 108k users who wrote those answers. The benchmarks below load the whole corpus. The workflow takes a median of 12.9 ms over 20 warm repetitions: 5.9 ms for the vector search, 5.0 ms for the SQL filter, and 2.0 ms for the Cypher traversal. Timings were measured on the same system as the comparison tables below.[1]

The Cypher step runs over a GAV, created as in the Property graph section, since the traversal is analytical. The graph benchmark below shows how much the GAV speeds it up.

## 5. Benchmarks

Is each model competitive enough for practical use, and what does it cost? We compare `arcadedb-embedded` against widely used Python-embeddable specialists for each model: SQLite and DuckDB (documents), LadybugDB (graph), and Chroma (vectors). “ArcadeDB” in the tables and text below means `arcadedb-embedded`, measured from Python.

**Protocol.** Every measurement runs in its own Docker container, one container at a time, with the same CPU cores and memory cap allocated to every backend, and is repeated 5 times in fresh containers. We report the median across the 5 repetitions with the full [min–max] range, since timing distributions are right-skewed and the median is the recommended summary [22], [23]. Engines run their shipped defaults except where stated.[2] In every table, higher is better for throughput and recall, and lower is better for everything else: latency, build and load times, memory, and disk.

The data is the Cross Validated (`stats.stackexchange.com`) public data dump [21], a statistics and machine-learning Q&A corpus of about 426k posts, 346k users, and 1.2M text embeddings. Embeddings are 384-dimensional `all-MiniLM-L6-v2` vectors over each question’s title and body. Vector-index parameters are matched across engines (degree as in the Vectors section, `ef_construction` $= 100$, `ef_search` $= 100$), and we report recall@10 against an exact ground truth. Graph OLAP for ArcadeDB uses a GAV. For document OLAP we first build a `NOTUNIQUE` index on each column the suite groups or filters by (`post_type`, `owner_user_id`, `score`), so the engine answers from an index instead of scanning the type; the index build is timed separately and excluded from query times. ArcadeDB and `arcadedb-embedded` are version 26.8.1 throughout.[3]

[1] Single host: 12th-gen Intel Core i9-12900HK (20 logical cores, eight allocated to each container), 61 GiB usable RAM, Samsung 980 PRO 2 TB NVMe SSD, Linux kernel 7.0.0 x86-64, Docker 29.5.3. Other versions: DuckDB 1.5.4, LadybugDB (`ladybug`) 0.18.1, Chroma 1.5.9, SQLite 3.46.1 as bound by Python’s `sqlite3` in `python:3.12-slim`. Embeddings are computed in the binding’s repository, L2-normalized, stored as `float32`.

[2] Each container is pinned to the same eight cores. SQLite runs WAL with `synchronous=NORMAL`, which its documentation recommends. ArcadeDB’s JVM heap is capped at 16 GiB, because the default 4 GiB does not complete the vector build; it is a ceiling, not a reservation, and the document workload peaks below 2 GiB under it. Peak memory is the container’s peak. Every result row records the versions and limits it ran under.

**Workloads.** Each model has a transactional (OLTP) and an analytical (OLAP) workload. OLTP is a mix of point operations issued by id, reported as throughput (ops/s). OLAP is a fixed suite of analytical queries, reported as the time to run the whole suite (ms).[4] The vector model instead runs 1,000 held-out nearest-neighbor queries and reports latency and recall@10. The exact query text for every workload is in the public benchmark suite.

**Documents (Table 2).** This workload runs the same SQL over the same posts that SQLite and DuckDB run over a table. SQLite is much faster than ArcadeDB on point operations, DuckDB is much faster on analytical queries, and ArcadeDB peaks at about 800 MiB of memory against about 300 MiB for the other two. Transactional throughput depends mostly on how often the engine syncs to disk. SQLite and ArcadeDB both run with relaxed durability here (SQLite's recommended write-ahead log (WAL) setting, ArcadeDB's default asynchronous flush), and at that setting SQLite does about 87,000 ops/s to ArcadeDB's 6,400. DuckDB syncs on every commit and has no setting to relax that, which is why it shows 219. With both SQLite and ArcadeDB forced to sync on every commit, they meet at the disk's limit, 262 against 187 ops/s.[5] On analytical SQL DuckDB answers the suite in about 9 ms versus ArcadeDB's 1,400 ms.

**Table 2.** *Document model (Cross Validated corpus): SQLite, DuckDB, ArcadeDB. OLTP is a mixed point read/insert/update workload (ops/s), OLAP an analytical aggregation suite (ms). SQLite and ArcadeDB run with relaxed durability (they do not sync to disk on every commit); DuckDB syncs on every commit. Median [min–max] over 5 reps. Peak = container memory, DB = on-disk size at the end of the OLTP rep (MiB). Best value per column in bold.*

| Backend | OLTP ops/s | OLAP ms | Ingest s | Peak MiB | DB MiB |
|---|---|---|---|---|---|
| SQLite | **87,150** [59,138–88,591] | 292.8 [284.4–293.9] | **0.33** [0.32–0.36] | **299** [277–304] | 20.1 |
| DuckDB | 219 [199–226] | **9.3** [9.3–9.6] | 0.38 [0.37–0.39] | 301 [296–308] | **17.3** |
| ArcadeDB | 6,416 [6,091–6,826] | 1,406.6 [1,382.1–1,432.4] | 17.52 [17.15–17.95] | 803 [763–846] | 38.5 |

**Graph (Table 3).** ArcadeDB answers point and one-hop reads faster than LadybugDB, LadybugDB is about 13× faster on the analytical suite, and ArcadeDB's graph costs far more memory and disk. On OLTP, ArcadeDB's 6× lead in throughput comes mostly from durability settings: LadybugDB syncs on every commit and has no setting to relax that. With ArcadeDB made to sync on every commit as well, it still leads, 595 against 525 ops/s. The OLTP reads do not depend on that setting, and there ArcadeDB is ahead at the median and at the 99th percentile (p99), though its single worst one-hop read is slower (Table 5).

**Table 3.** *Graph model (Cross Validated corpus): LadybugDB, ArcadeDB. OLTP is neighborhood/traversal point ops (ops/s), OLAP a multi-query analytical suite (ms), Ingest the bulk load of vertices and edges (s).*

[3] Single host: 12th-gen Intel Core i9-12900HK (20 logical cores, eight allocated to each container), 61 GiB usable RAM, Samsung 980 PRO 2 TB NVMe SSD, Linux kernel 7.0.0 x86-64, Docker 29.5.3. Other versions: DuckDB 1.5.4, LadybugDB (`ladybug`) 0.18.1, Chroma 1.5.9, SQLite 3.46.1 as bound by Python's `sqlite3` in `python:3.12-slim`. Embeddings are computed in the binding's repository, L2-normalized, stored as `float32`.

[4] Document OLTP issues 5,000 operations (60% reads, 20% updates, 10% inserts, 10% deletes); graph OLTP issues 2,000 (50% point lookups, 35% one-hop traversals, 15% vertex inserts). The document OLAP suite is five `GROUP BY`/aggregate/top-N queries over the posts table; the graph suite is four traversal-and-aggregation queries (e.g. top contributors by post count, multi-hop path counts). Each query's time is the average of seven runs, and the suite time is their sum.

[5] `arcadedb.txWalFlush=2`, and SQLite at its library defaults of a rollback journal with `synchronous=FULL`. The SQLite strict figure is in the suite's append log rather than the frozen results table.

*ArcadeDB OLAP uses a GAV, whose build time is shown separately. LadybugDB syncs to disk on every commit; ArcadeDB runs at its relaxed default. Median [min–max] over 5 reps. Peak = container memory, DB = on-disk size at the end of the OLTP rep (MiB). Best value per column in bold.*

| Backend | OLTP ops/s | OLAP ms | GAV build s | Ingest s | Peak MiB | DB MiB |
|---|---|---|---|---|---|---|
| LadybugDB | 525 [467–532] | **65.7** [64.9–66.9] | n/a | **0.63** [0.61–0.64] | **684** [675–688] | **41.4** |
| ArcadeDB | **3,212** [3,034–3,629] | 839.6 [794.7–875.1] | 1.49 [1.39–1.77] | 26.67 [26.21–26.92] | 11,037 [10,561–11,190] | 1,840.8 |

On graph analytics LadybugDB wins, 66 ms against 840 ms. The GAV is what gets ArcadeDB to 840 ms: without it the same suite takes 2,023 ms, against a 1.5 s build. It does not change transactional throughput. The graph also costs far more space: 1,841 MiB on disk against LadybugDB's 41 MiB, and about 11 GiB of peak memory against 684 MiB.

**Vectors (Table 4).** At matched graph degree, ArcadeDB's recall@10 is slightly higher than Chroma's (0.980 against 0.972) and its peak memory is 14% lower, because it keeps the vectors on disk while Chroma's index holds them all in RAM. Chroma builds the index in about 60% of the time and answers queries about 3.5× faster, 1.2 ms against 4.2 ms, both single-digit milliseconds.

**Table 4.** *Vector model (Cross Validated corpus, 1.2M vectors): Chroma, ArcadeDB at matched HNSW degree (Chroma $M = 16$, ArcadeDB `maxConnections`=32, both `ef_construction`=100, `ef_search`=100). Build = insert+index (s), Query = mean latency per query in a rep (ms), recall@10 vs exact ground truth. Median [min–max] over 5 reps. Peak = container memory, DB = on-disk size (MiB). Best value per column in bold.*

| Backend | Build s | Query ms | recall@10 | Peak MiB | DB MiB |
|---|---|---|---|---|---|
| Chroma | **319.1** [315.7–320.6] | **1.17** [1.15–1.17] | 0.972 [0.970–0.972] | 25,212 [25,173–27,002] | **2,208** |
| ArcadeDB | 546.1 [541.3–550.6] | 4.15 [3.86–4.17] | **0.980** [0.978–0.981] | **21,650** [19,680–21,791] | 2,856 |

Starting the JVM and opening the database together take about 0.3 s, once per process, less than Chroma's Python import alone (0.37 s).

**Latency (Table 5).** On the transactional reads and the vector queries, ArcadeDB's median and p99 latencies beat DuckDB and LadybugDB and trail SQLite and Chroma. Its worst single operation, though, is 35 to 146 ms across the three models against 0.05 to 11 ms for the specialists. For most applications this rare slow request does not matter, but it does if every request has a strict deadline.

**Table 5.** *Per-operation latency (ms) of the read operations in the OLTP workloads and of the vector queries, on the Cross Validated corpus: the median operation, the p99 tail, and the single worst operation, each reported as the median over 5 repetitions. Best value per group in bold.*

| Model / op | Backend | median | p99 | max |
|---|---|---|---|---|
| document read | SQLite | **0.004** | **0.007** | **0.05** |
| | DuckDB | 0.92 | 1.89 | 2.9 |
| | ArcadeDB | 0.06 | 0.17 | 34.9 |
| graph point | LadybugDB | 0.41 | 1.24 | 1.8 |
| | ArcadeDB | **0.18** | **0.74** | **1.5** |
| graph hop | LadybugDB | 1.44 | 4.34 | **11.0** |
| | ArcadeDB | **0.21** | **0.81** | 90.3 |
| vector query | Chroma | **1.18** | **1.36** | **1.4** |
| | ArcadeDB | 3.86 | 7.11 | 145.9 |

**Memory.** On the document workload ArcadeDB peaks at about 800 MiB against about 300 MiB for SQLite and DuckDB. The difference is the cost of running a JVM, with the engine's heap and page cache inside it. The graph build is the one place it is far larger, and on vectors its disk-backed index uses less memory than Chroma's.

### *5.1. Result transport into Python*

The same engine runs the same query for all four methods, and only the way results cross into Python differs. `iter_dicts()` converts one row at a time, `to_json_list()` serializes one JSON string per batch for Python to parse, `to_columns()` reads a packed columnar buffer with `numpy.frombuffer()`, and `to_arrow()` wraps that same buffer as a `pyarrow.Table`.

**Table 6**. *Result transport, 200k-row document type, four columns (two integer, one double, one string). Median of 7 timed passes after 2 warmups, milliseconds. The engine executes the same query for every method.*

| Rows | `iter_dicts()` | `to_json_list()` | `to_columns()` | `to_arrow()` |
|---|---|---|---|---|
| 10 | 2.04 | 1.70 | 0.82 | 0.59 |
| 100 | 2.66 | 1.10 | 0.93 | 0.79 |
| 1,000 | 20.12 | 2.97 | 2.28 | 1.49 |
| 10,000 | 197.29 | 25.87 | 15.87 | 14.20 |
| 100,000 | 1,978.27 | 278.22 | 135.51 | 109.03 |

At 100k rows (Table 6) `iter_dicts()` is 18$\times$ slower than `to_arrow()` (1,978 against 109 ms), and the whole difference is on the Python side. At 10 rows every method finishes under 2.1 ms, so the choice only starts to matter as results grow.

`to_arrow()` is the fastest method at every size; at 100k rows it is about 20% faster than `to_columns()`. The difference is the string column: the NumPy path decodes it into one Python `str` per row, while Arrow keeps the string layout the buffer already has. A purely numeric result would narrow this.

`arcadedb-embedded` makes the ArcadeDB Java engine usable as an embedded database from Python. Python is a thin layer over the engine, and the layer does add overhead, but it is small: compared with a pure Java program running the same vector search on the same engine, going through Python adds about 10% to the latency. The query itself does the same work in both. The overhead only becomes visible on large results, and the transport table above shows how much the method chosen to return them matters.

## 6. When to use it

`arcadedb-embedded` is most useful when an application needs two or more of documents, a graph, and vector search over the same data, on one machine, and wants them in one process with one set of identifiers and one transaction scope. The typical case is the hybrid workflow shown above, where one query has to combine a similarity search with filters over documents and a traversal over the graph. In that setting it is fast enough: point reads and graph traversals answer in well under a millisecond, vector queries in single-digit milliseconds at the same recall as a dedicated index, and everything is ACID.

The trade-off is that each specialist is faster on its own model: DuckDB by two orders of magnitude on analytical SQL, LadybugDB by an order of magnitude on graph analytics,

SQLite on point reads and writes of documents, and Chroma on building and querying a vector index. For a workload that lives entirely in one model, the specialist is the natural choice. Covering all three has a price in memory and loading time, as the benchmarks show.

## 7. Conclusion

Applications that need documents, graph, and vectors over the same data usually run three systems and application code to keep them consistent. `arcadedb-embedded` packages the ArcadeDB Java engine as a Python wheel with its own JRE, so one engine provides documents, graph, and vectors from one process. The hybrid workflow uses all three in turn, a vector search, then a SQL filter over the documents, then a graph traversal, without data leaving the engine. In the benchmarks each specialist is faster on its own model, but `arcadedb-embedded` holds its own while covering all three: it beats LadybugDB on transactional graph work, beats DuckDB on point reads, and matches Chroma's recall with less memory. Being a generalist has a price, mainly in memory and loading time. Both ArcadeDB and `arcadedb-embedded` release monthly and are under active development, so these numbers describe one release, 26.8.1, and later releases are expected to improve on them. The ArcadeDB project wrote the engine; this paper contributes the binding, the packaging, and the in-process workflow. The binding (Apache-2.0), the benchmark suite, and the results behind every table are public:

- Repository: https://github.com/humemai/arcadedb-embedded-python
- Documentation: https://docs.humem.ai/arcadedb/
- Benchmark suite and results: https://github.com/humemai/arcadedb-embedded-python/tree/main/benchmarks/python-bindings

### Acknowledgements

We thank the ArcadeDB open-source contributors for the engine that this work makes available to Python, and the JPype maintainers for the CPython–JVM bridge that the binding builds on.

Claude (Anthropic) was used to draft and revise the text and to help carry out the benchmark experiments. The author reviewed and verified all of it and is responsible for the final content.

### References

[1] M. Raasveldt and H. Mühleisen, "DuckDB: an Embeddable Analytical Database," in *Proceedings of the 2019 International Conference on Management of Data (SIGMOD '19)*, New York, NY, USA, 2019, pp. 1981–1984. doi: 10.1145/3299869.3320212.

[2] SQLite Development Team, "SQLite." [Online]. Available: https://www.sqlite.org/

[3] A. A. Hagberg, D. A. Schult, and P. J. Swart, "Exploring Network Structure, Dynamics, and Function using NetworkX," in *Proceedings of the 7th Python in Science Conference (SciPy 2008)*, G. Varoquaux, T. Vaught, and J. Millman, Eds., Pasadena, CA, USA, 2008, pp. 11–15. doi: 10.25080/TCWV9851.

[4] LadybugDB contributors, "LadybugDB: an embedded graph database (Python package ladybug)." [Online]. Available: https://github.com/LadybugDB/ladybug

[5] M. Douze *et al.*, "The Faiss Library," *IEEE Transactions on Big Data*, vol. 12, no. 2, pp. 346–361, 2026, doi: 10.1109/TBDATA.2025.3618474.

[6] Y. Malkov and hnswlib contributors, "hnswlib: Header-only C++/Python library for fast approximate nearest neighbors." [Online]. Available: https://github.com/nmslib/hnswlib

[7] Chroma, "Chroma: the open-source data infrastructure for AI." [Online]. Available: https://github.com/chroma-core/chroma

[8] D. Datta, "Python is all you need: an overview of the composable, Python-native data stack," in *Proc. 24th Python in Science Conf. (SciPy 2025)*, 2025, pp. 75–84. doi: 10.25080/wjjm7869.

[9] L. Garulli, “ArcadeDB.” [Online]. Available: https://arcadedb.com/

[10] K. E. Nelson and M. K. Scherer, “JPype.” [Online]. Available: https://www.osti.gov/doecode/biblio/46107

[11] T. Kim and The ArcadeDB contributors, “ArcadeDB Embedded for Python.” Zenodo, 2026. doi: 10.5281/zenodo.21781540.

[12] A. Kane, “pgvector: Open-source vector similarity search for Postgres.” [Online]. Available: https://github.com/pgvector/pgvector

[13] Neo4j, Inc., “Neo4j Graph Database.” [Online]. Available: https://neo4j.com/

[14] J. Wang *et al.*, “Milvus: A Purpose-Built Vector Data Management System,” in *Proceedings of the 2021 International Conference on Management of Data (SIGMOD '21)*, 2021, pp. 2614–2627. doi: 10.1145/3448016.3457550.

[15] M. Raasveldt and H. Mühleisen, “DuckDB.” [Online]. Available: https://github.com/duckdb/duckdb

[16] Y. A. Malkov and D. A. Yashunin, “Efficient and Robust Approximate Nearest Neighbor Search Using Hierarchical Navigable Small World Graphs,” *IEEE Transactions on Pattern Analysis and Machine Intelligence*, vol. 42, no. 4, pp. 824–836, 2020, doi: 10.1109/TPAMI.2018.2889473.

[17] DataStax and JVector contributors, “JVector: the most advanced embedded vector search engine.” [Online]. Available: https://github.com/datastax/jvector

[18] S. J. Subramanya, F. Devvrit, H. V. Simhadri, R. Krishnaswamy, and R. Kadekodi, “DiskANN: Fast Accurate Billion-point Nearest Neighbor Search on a Single Node,” in *Advances in Neural Information Processing Systems*, 2019. [Online]. Available: https://proceedings.neurips.cc/paper_files/paper/2019/hash/09853c7fb1d3f8ee67a61b6bf4a7f8e6-Abstract.html

[19] H. Schreiner, J.-C. Fillion-Robin, and M. McCormick, “Scikit-build-core: A modern build-backend for CPython C/C++/Fortran/Cython extensions,” in *Proceedings of the 23rd Python in Science Conference (SciPy 2024)*, 2024, pp. 225–235. doi: 10.25080/FMKR8387.

[20] J. Laskey, “JEP 282: jlink: The Java Linker.” [Online]. Available: https://openjdk.org/jeps/282

[21] Stack Exchange Inc., “Cross Validated (stats.stackexchange.com) data dump.” Internet Archive, https://archive.org/details/stackexchange, 2024.

[22] M. Raasveldt, P. Holanda, T. Gubner, and H. Mühleisen, “Fair Benchmarking Considered Difficult: Common Pitfalls In Database Performance Testing,” in *Proceedings of the Workshop on Testing Database Systems (DBTest)*, 2018. doi: 10.1145/3209950.3209955.

[23] T. Hoefler and R. Belli, “Scientific Benchmarking of Parallel Computing Systems: Twelve Ways to Tell the Masses when Reporting Performance Results,” in *Proceedings of the International Conference for High Performance Computing, Networking, Storage and Analysis (SC)*, 2015. doi: 10.1145/2807591.2807644.